\documentclass[oneside,onecolumn]{article}

\usepackage{bm}         

\usepackage{amssymb}    
\usepackage{amsfonts}   

\mathchardef\ordinarycolon\mathcode`\:                     %
\def\vcentcolon{\mathrel{\mathop\ordinarycolon}}           
\providecommand*\coloneqq{{\;\vcentcolon\mkern-0.8mu=\;}}  

\begin{document}

\title{\bf Einstein's Real ``Biggest Blunder''}
\bigskip\bigskip

\author{Homer G.~Ellis \\ \\
Department of Mathematics \\
University of Colorado Boulder \\
Boulder, Colorado 80309--0395 \\
Email: ellis@euclid.colorado.edu}

\maketitle

\noindent
ABSTRACT. $\;$ Albert Einstein's real ``biggest blunder'' was not the 1917
introduction into his gravitational field equations of a cosmological constant
term $\Lambda$, rather was his failure in 1916 to distinguish between the
entirely different concepts of {\it active} gravitational mass and
{\it passive} gravitational mass.  Had he made the distinction, and followed
David Hilbert's lead in deriving field equations from a variational principle,
he might have discovered a true (not a cut and paste) Einstein-Rosen bridge and
a cosmological model that would have allowed him to predict, long before such
phenomena were imagined by others, inflation, a big bounce (not a big bang), an
accelerating expansion of the universe, dark matter, and the existence of
cosmic voids, walls, filaments, and nodes.

\vskip 20pt

Albert Einstein is reputed to have said that his ``biggest blunder'' was the
1917 introduction into his gravitational field equations of a cosmological
constant term $\Lambda$~\cite{eins1,gamo}.  In retrospect this looks to have
been closer to an unwitting act of prescience than to a blunder, as $\Lambda$
seems to offer a route to understanding (in a superficial way) the acceleration
of the expansion of the universe.

What has gone unrecognized for almost a century is that already in 1916
Einstein had made a {\it real} blunder by failing to distinguish between the
entirely different concepts of {\it active} gravitational mass and
{\it passive} gravitational mass.  That he confused the two becomes clear upon
a careful reading of \S 16 of his paper Die Grundlage der allgemeinen
Relativit\"atstheorie~\cite{eins2}, in which he sought to extend his tensorial
field equations for empty space to the case in which space is permeated by a
continuous distribution of gravitating matter, in analogy to the extension of
the Laplace equation $\nabla^2 \phi = 0$ for the vacuum gravitational potential
$\phi$ of newtonian gravity to the Poisson equation
$\nabla^2 \phi = 4 \pi \kappa \mu$, where $\kappa$ is Newton's gravitational
constant and $\mu$ denotes, in Einstein's imprecise choice of words, the
``density of matter''.  Whatever he had in mind when he wrote that phrase, it
is clear that $\mu$ is the active (gravitat{\it ing}), not the passive
(gravitat{\it ed}), mass density of the matter in question.

The (weak) principle of equivalence, which Einstein had specifically built into his general theory of relativity, identifies passive gravitational mass with
inertial mass, and the special theory identifies inertial mass with energy via
$E = m c^2$.  When Einstein, continuing in \S 16, wrote that (in the special
theory) ``energy\ldots finds its complete mathematical expression in a tensor
of the second rank, the energy-tensor'', and then, relying on these
identifications, proceeded to introduce as the material source term in his
field equations ``a corresponding energy-tensor of matter
$\rm{T}^\alpha_\sigma$'', he implicitly assumed, without presenting any
justification for it, that {\it passive} gravitational mass could play the same
role in his equations that {\it active} gravitational mass plays in the Poisson
equation of newtonian gravity.  That was his real ``biggest blunder''.  

It is a reasonable speculation that Einstein was seduced into this logical
error by the fact that in newtonian gravity application of the law of action
and reaction to the forces exerted on each other by two gravitating bodies A
and B allows the inference that the ratio of active mass to passive mass is the
same for B as it is for A, thus by extension is the same for all such material
bodies.  If he was guided by this, whether consciously or unconsciously, then
he was misguided, for this application of the action-reaction law to bodies not
in contact requires the assumption that gravity acts instantaneously over the
intervening distance, an assumption at odds with the finiteness of the
propagation velocity of gravitational effects implicit in Einstein's theory.

Whatever caused Einstein to make the error, it was solidified in his mind by a
consequence he asserted to be ``the strongest reason for the choice'' of his
``energy-tensor'' source term for a ``frictionless adiabatic fluid'' of
``density''~$\rho$, pressure~$p$, and proper four-velocity
distribution~$u^\alpha$.  The choice he made was
$T^{\alpha \beta} = \rho u^\alpha u^\beta - (p/c^2) g^{\alpha \beta}$
(revised to
$T^{\alpha \beta} = (\rho + p) u^\alpha u^\beta - (p/c^2) g^{\alpha \beta}$ by
his redefintion $\rho \to \rho + p$).  The consequence he liked was that the
vanishing of the divergence of the left member of the resulting field equation
{${\bm R}^{\alpha \beta} - \frac12 {\bm R} g^{\alpha \beta}
= \frac{8 \pi \kappa}{c^2} T^{\alpha \beta}$}
(in today's notation) implied the vanishing of the divergence
${T^{\alpha \beta}}_{: \beta}$ of $T^{\alpha \beta}$, which, interpreted as
``the Eulerian hydrodynamical equations of the general theory of
relativity'', \ldots ``give a complete solution of the problem of
motion''~(\cite{eins2},\S 19).  That would certainly be an alluring consequence
if it were not based on the confusion between the inertial-passive mass density
$\rho$ and the active gravitational mass density $\mu$ of the fluid in
question.

Had Einstein recognized the error implicit in making the ``energy-tensor of
matter $\rm{T}^\alpha_\sigma$'' the material source term in his field
equations, what road might he then have followed?  Having already incorporated
into his theory a variational principle to identify worldlines of test
particles as geodesic paths of the space-time metric, he might have looked for
a variational principle from which to derive field equations.  That is what
David Hilbert did contemporaneously to derive field equations with the
electromagnetic field providing the source term (field equations which for the
vacuum matched those of Einstein, giving rise to a still ongoing debate over
assignment of priority)~\cite{hilb}.  What neither Einstein nor Hilbert thought
to do was look for inspiration to the variational principle from which the
Poisson equation for $\phi$ derives, namely,
\begin{equation}
\delta \! \int (|\nabla \phi|^2 + 8 \pi \kappa \mu \phi) \, d^3\!x = 0 \, .
\label{eqn1}
\end{equation}
From Hilbert's variational principle it would have been straightforward to
arrive at the generalization 
\begin{equation}
\delta \! \int \left({\bm R} - {\frac{8 \pi \kappa}{\rm c^2}} \mu \right)
               \, |g|^{\frac12} \, d^4\!x = 0 
\label{eqn2}
\end{equation}
of~(\ref{eqn1}),
for which the Euler--Lagrange equations are equivalent to
\begin{equation}
{\bm R}^{\alpha \beta} - {\textstyle\frac12} {\bm R} \, g^{\alpha \beta}
 = \displaystyle{-\frac{4 \pi \kappa}{\rm c^2}} \mu g^{\alpha \beta} \, ,
\label{eqn3}
\end{equation}
equivalent in turn to
${\bm R}_{\alpha \beta} = \frac{4 \pi \kappa}{c^2} \mu g_{\alpha \beta}$, the
00 component of which reduces in the slowly varying, weak field approximation
to the Poisson equation, with $\phi = \frac12 (g_{00} - c^2)$.
This would have had the salutary effect of replacing Einstein's unjustified
``energy-tensor'' source term by 
$T^{\alpha \beta} = -\frac{4 \pi \kappa}{\rm c^2} \mu g^{\alpha \beta}$,
properly built with $\mu$ instead of $\rho$.  A not so salutary effect is that
the vanishing of the divergerce of $T^{\alpha \beta}$ would imply that $\mu$
must be constant, which would suggest that the only cosmological models that
could be solutions of~(\ref{eqn3}) would be the discredited `steady-state'
models.  A way out of this dilemma will present itself below.

Let us suppose that Einstein had arrived at~(\ref{eqn3}) by way of the
variational principle~(\ref{eqn2}), and published that instead of the equations
he did publish.  How might subsequent developments in the general theory of
relativity have played out?  No doubt Schwarzschild would have derived the
same well-known blackhole metric he published in 1916, for the vacuum equations
are the same, but when he searched for an interior solution to go with it he
would have found a model for the gravitational field inside a star simpler and
more understandable than the one he published as his `interior solution' of
Einstein's original equations~\cite{elli1}.  What else?

We know that in 1935 Einstein and Rosen tried to construct a singularity-free
model of an elementary particle as a `bridge' (later termed a `wormhole')
connecting two copies of the region of Schwarzschild's blackhole outside the
$r = 2 m$ horizon, but could do so only by weakening the field equations in a
mathematically suspect manner (replacing ${\bm R}_{\alpha \beta}$ with
$g^2 {\bm R}_{\alpha \beta}$, where $g$ is the determinant of the metric
tensor)~\cite{eiro}.  Because their method would not work if $m$ were negative,
they believed that they had found a reason why such a particle could have only
a positive or a zero inertial mass, once again confusing active gravitational
mass (the $m$ of the Schwarzschild blackhole) with the conceptually distinct
inertial-passive mass.  Now let us imagine that Einstein (with or without
Rosen), instead of contenting himself with this {\it ad hoc} cut and paste job,
had investigated Schwarzschild's solution more thoroughly and had come across
one of the papers by Gullstrand~\cite{gull} and Painlev\'e~\cite{pain} that
expressed Schwarzschild's blackhole metric in the form
$dt^2 - (dr - v \, dt)^2 - r^2 d\Omega^2$, where $v = -\sqrt{\frac{2m}{r}}$,
the velocity of an observer free-falling from rest at $r = \infty$.  He might
then have noticed that to such an observer space, in the sense of a
$t$ = constant slice of space-time, would have no curvature, its metric being
that of euclidean three-space in spherical coordinates, namely,
$dr^2 + r^2 d\Omega^2$.  Studying the matter further he could have realized
that such curvature would be admitted if the field equations were weakened by
intoduction of a scalar field $\psi$ minimally coupled to the space-time
geometry.  This would be accomplished by modifying the integrand of the action
integral of~(\ref{eqn2}) to
${\bm R} - \frac{8 \pi \kappa}{c^2} \mu -
2 \psi^{.\gamma} \psi_{.\gamma}$.  But this new, enlightened Einstein would not
stop there, for having recognized that `energy', if coupled to geometry at all,
need not be coupled with the same polarity that ordinary matter is coupled
with, and, seeing no reason to choose one polarity over the other, would use a
second scalar field $\phi$ (not the newtonian $\phi$) to further modify the
integrand to
${\bm R} - \frac{8 \pi \kappa}{c^2} \mu +
2 \phi^{.\gamma} \phi_{.\gamma} - 2 \psi^{.\gamma} \psi_{.\gamma}$.
Moreover, the new Einstein, holding just as the old to the principle that all
things physical were to be found embedded intrinsically in the geometry of
space-time, would derive new field equations from the thus modified variational
principle by varying only the metric components, and not the scalar fields (or
for that matter, the density $\mu$).  The field equations that would have
resulted are
\begin{equation}
{\bm R}_{\alpha \beta} - {\textstyle\frac12} {\bm R} \, g_{\alpha \beta}
  = -\frac{4 \pi \kappa}{c^2} \mu  \, g_{\alpha \beta}
     - 2 \, (\phi_{.\alpha} \phi_{.\beta}
     - {\textstyle\frac12} \phi^{.\gamma} \phi_{.\gamma}
       \, g_{\alpha \beta})
     + 2 \, (\psi_{.\alpha} \psi_{.\beta}
     - {\textstyle\frac12} \psi^{.\gamma} \psi_{.\gamma}
       \, g_{\alpha \beta}) \, .
\label{eqn4}
\end{equation} 

Pressing on, Einstein (perhaps with the help of Rosen) might well have arrived
in 1935 at the true Einstein-Rosen bridge first discovered around 1970
independently of one another by two researchers, and called descriptively by
one of them a `drainhole' with a gravitational `ether' flowing through
it~\cite{elli2,bron}.  This geodesically complete, nonsimply connected
space-time manifold lacks both the singularity and the event horizon of the
Schwarzschild blackhole.  It is specified by assignment of two parameters:
$m$, the active gravitational mass of the drainhole as measured at infinity on
the `high' side, and $n$, which determines the size of the drainhole, subject
only to the constraints that $0 \leq m < n$.  Had he dug far enough, Einstein
would have found that, because the `ether', describable as a cloud of test
particles free-falling from rest at infinity on the high side (or simply
as space itself flowing from the high side to the low side), accelerates
downward all the way into and through the drainhole and out to minus infinity
on the low side, the drainhole gravitationally attracts all matter on its high
side but {\it repels} all matter on the low side.  Digging even deeper he would
have learned that the negative mass parameter $\bar m$ specifying the strength
of the repulsion on the low side exceeds in magnitude the attractive mass
parameter $m$ in the ratio $-\bar m/m = e^{m \pi/\sqrt{n^2 - m^2}}$.  An
inevitable conclusion from this would be that in addition to the positive mass
density $\mu$ there must exist a negative mass density $\bar \mu$, thus an
overall negative net active gravitational mass density $\mu + \bar \mu$ to
replace the $\mu$ in~(\ref{eqn4}), turning it into
\begin{equation}
{\bm R}_{\alpha \beta} - {\textstyle\frac12} {\bm R} \, g_{\alpha \beta}
  = -\frac{4 \pi \kappa}{c^2} (\mu + \bar \mu)  \, g_{\alpha \beta}
     - 2 \, (\phi_{.\alpha} \phi_{.\beta}
     - {\textstyle\frac12} \phi^{.\gamma} \phi_{.\gamma}
       \, g_{\alpha \beta})
     + 2 \, (\psi_{.\alpha} \psi_{.\beta}
     - {\textstyle\frac12} \psi^{.\gamma} \psi_{.\gamma}
       \, g_{\alpha \beta}) \, .
\label{eqn5}
\end{equation} 
Having gone this far, the new Einstein would surely have `divergenced' this
equation and arrived at
\begin{equation}
2 \, (\square \phi) \phi_{.\alpha} - 2 \, (\square \psi) \psi_{.\alpha}
  \coloneqq 2 \, \phi^{.\gamma}\!{}_{:\gamma} \phi_{.\alpha}
            - 2 \, \psi^{.\gamma}\!{}_{:\gamma} \psi_{.\alpha}
  = -\frac{4 \pi \kappa}{c^2} (\mu + \bar \mu)_{.\alpha} \, ,
\label{eqn6}
\end{equation} 
which he could have recognized as a weaker replacement for the equations
$\square \phi = 0$ and $\square \psi = 0$ that varying $\phi$ and $\psi$
would have produced, and which would have required neither $\mu$, nor
$\bar \mu$, nor their net $\mu + \bar \mu$ to be constant.

The new, enlightened Einstein might well have introduced in 1917 a cosmological
constant $\Lambda$, for the same reasons that the old Einstein relied on, and
after Hubble's discovery that the universe is expanding, consigned it to the
trash bin just as before.  But after 1935 he might well also have revisited the
question, seen that his cosmological constant was in effect the negative net
density $\mu + \bar \mu$ in disguise, and retracted his characterization of it
as his ``biggest blunder'' (thus reducing his ``biggest blunder'' count to 
zero).  Investigating the cosmological consequences of~(\ref{eqn5}) he might
easily have discovered the model of the universe presented in~\cite{elli3}, and
thereby have {\it predicted}, long before such phenomena were imagined by
others, inflation, a big bounce (not a big bang), an accelerating expansion of
the universe, dark matter, and the existence of cosmic voids, walls, filaments,
and nodes.

All of that is only a small part of what might have transpired had Einstein
recognized his real ``biggest blunder'' and followed the road that then would
have stretched out before him.  For those who admire and respect the work that
he did it should be a cause for real regret that in the long and arduous
process of constructing his general theory of relativity he missed this
opportunity to make the theory logically consistent and therefore more
powerful.

\end{document}